\documentclass[mathleft
]{an}
\usepackage{graphicx}
\usepackage{times}
\begin{document}

\Pagespan{789}{}
\Yearpublication{2006}%
\Yearsubmission{2005}%
\Month{11}%
\Volume{999}%
\Issue{88}%

\title{Signature of a magnetic activity cycle in HD49933 observed by CoRoT}

\author{R.A. Garc\'\i a\inst{1}\fnmsep\thanks{Corresponding author:
  \email{rgarcia@cea.fr}\newline}
\and J. Ballot\inst{2} 
\and S. Mathur\inst{3}
\and D. Salabert\inst{4,5}
\and C. R\'egulo\inst{4,5}
}
\titlerunning{Signature of a magnetic activity cycle in HD49933 observed by CoRoT}
\authorrunning{Garc\'\i a et al.}
\institute{
Laboratoire AIM, CEA/DSM-CNRS, Universit\'e Paris 7 Diderot, IRFU/SAp-SEDI, Centre de Saclay, 91191, Gif-sur-Yvette, France
\and 
Laboratoire d'Astrophysique de Toulouse-Tarbes, Universit\'e de Toulouse, CNRS, F-31400, Toulouse, France
\and 
High Altitude Observatory, NCAR, P.O. Box 3000, Boulder, CO 80307, USA
\and
 Universidad de La Laguna, Dpto de Astrof\'isica, 38206, Tenerife, Spain
 \and 
 Instituto de Astrof\'\i sica de Canarias, 38205, La Laguna, Tenerife, Spain
}

\received{}
\accepted{}
\publonline{later}


\abstract{Solar-like stars with an external convective envelope can develop magnetic activity cycles under the interaction of convection, rotation and magnetic fields. Even in the Sun, these dynamo effects are not yet well understood and it would be extremely important to extend this study to different stars with different characteristics. HD49933 is a F5V, 1.2 solar mass star that has been observed by CoRoT for 60 days during the initial Run and 137 more days about 6 months later. Thus, a total of 400 days have been covered with the two observations. Assuming that the activity cycle is proportional to the rotation of the star (which spins 8-9 times faster than the Sun, i.e., $P_{rot}$=3.4 days), CoRoT observations could be good to perform such study. The analysis techniques employed here hve already been successfully tested on sun-as-a-star observations done by VIRGO and GOLF on board SoHO and ground-based MARK-I instrument.
}

\maketitle

\section{Introduction}
Astronomers of modern era have been observing the activity on the surface of the Sun since Galileo and they have been tracking the number of sunspots since then. The 11-year cycle of the observed active regions on the solar surface is a consequence of a dynamo process running in the convective region of the Sun (Dikpati \& Gilman 2006). However, the delayed onset of solar cycle 24 (e.g. Salabert et al. 2009) gives evidence of a lack of understanding the physical processes governing the activity cycle. Thus, observing such activity cycles in other stars would, in principle, help to improve our knowledge in this field. 

Stellar activity of hundreds of stars have already been monitored over the last 40 years with the Mount Wilson survey (Wilson 1978; Baliunas \& Vaughan 1985; Baliunas et al. 1995) or the Lowell Observatory survey (Hall, Lockwood \& Skiff 2007). Many stars seem to have periodic cycles between 2.5 and 25 years. Moreover, it has been suggested that there are two different branches of stellar cycles in solar-type stars, one active and one inactive (Saar \& Brandenburg 1999), which could reflect the existence of two different dynamos operating in their interiors.

To properly study the dynamo models, we need to complement the classical picture of the stars with asteroseismic informations (e.g. Stello et al. 2009). This will not only give access to complementary information on the activity cycle --due to the changes in the oscillation mode frequencies and amplitudes-- but also provide direct information of the structure and dynamics of the stellar interior. In particular, it would allow us to study the extension of the convective zone (e.g. Ballot, Turck-Chi\'eze \& Garc\'\i a 2004), the differential rotation, and distinguish between surface rotation rate and the stellar inclination axis whenever possible (see the discussion on Gizon \& Solanki (2003) and Ballot, Garc\'\i a \& Lambert (2006)).

In the Sun, the central frequency of the acoustic modes (p modes) changes between the maximum and the minimum of the solar cycle: the modes are shifted by $\approx$0.45 $\mu$Hz towards higher frequencies at the maximum of the magnetic activity (e.g. Chaplin et al. 2007). Indeed, not only the frequencies, but all the characteristics of the p modes change with the activity cycle (e.g. Jim\'enez-Reyes, et al. 2007). In particular, these frequency shifts --during a maximum of activity-- are also accompanied in the Sun by a reduction of the observed amplitude of the modes in both: velocity and intensity fluctuations (e.g. JimŽnez-Reyes et al. 2003).

For a cool star, like the Sun, with an alpha-Omega dynamo, a longer period rotation implies a longer cycle period (e.g. Thomas \& Weiss 2008):
\begin{equation}
  \label{eq1}
 \frac{P_{cyc}}{P_{rot}} = \frac{\Omega}{\Omega_{cyc}} = CR_o^{q}
 \end{equation}
with $R_o$=$P_{rot}/\tau_c$, the Rosby number and $\tau_c$ the convective turnover time and q changing from 0.25 to 1 (e.g. Ossendrijver 1997; Saar 2002; Jouve, Brown \& Brun 2010). Using $\tau_c\approx 12$d and following the computations done by Kim \& Demarque (1996), we find that $R_o/R_{o\sun}\approx 0.4$, hence $P_{cyc}\approx$ 200 -- 400 days.

The NASA Kepler mission (Koch et al. 2010) is going to provide a unique opportunity to do asteroseismic measurements of the same field during more than 3.5 years of solar-like stars (Chaplin et al 2010), red giants (Bedding et al. 2010), stars belonging to open clusters (Stello et al. 2010), and even discover new hybrid pulsators (Grigahcene et al. 2010). In all cases we will be able to study long period activity cycles (e.g. Karoff et al. 2009) as well as their seismic properties. This was impossible from ground due to the shortness of the observational campaigns (e.g. Arentoft et al. 2008).  We already have more than 3 years of data coming from the CoRoT mission (e.g. Michel et al. 2008) in which several solar-like stars have been observed (Barban et al. 2009; Garc\'\i a et al. 2009; Mosser et al. 2009; Deheuvels et al. 2010; Mathur et al. 2010a). Unfortunately, the longest observations available are limited to around 150 days. There is one exception, the solar-like star HD49933. Indeed,  assuming that magnetic activity cycles in stars are proportional to its rotation, HD49933 could be a good target to seismically study the existence of a stellar activity cycle. Its surface spins around 8-9 times faster than the Sun (Appourchaux et al. 2008; Benomar et al. 2009) which means that the time span of 400 days covered by the two observation runs could cover a significant part of such a magnetic cycle (Garc\'\i a et al. 2010). This star was indeed known to be active (Mosser et al. 2004).

\section{Methodology and data analysis}
To check whether or not there is a magnetic activity cycle in HD49933 we have performed three different, but complementary studies: a) study the evolution of the surface rotation peak in the PSD, b) look for a frequency shift in the p modes, and c) study the rms maximum amplitude per radial mode, $A_{max}$.

We have analyzed the so-called `"Helreg'' (Auvergne et al. 2009) light curves of HD49933 provided by the CoRoT satellite during $\sim$60 days of the initial run in 2007 (Appourchaux et al. 2008) and during $\sim$137 more days in 2008 separated about 6 months (Benomar et al. 2009), covering a total of 400 days.

\section{Time evolution of the active regions}

The passage of the active regions crossing the visible stellar disk is a very good indicator of the activity. Mathur et al. (2008) used this methodology to study the activity cycle of the Sun using helioseismic radial velocity observations of the GOLF/SoHO instrument (Gabriel et al. 1995; Garc\'\i a et al. 2005) compared to 5 more classical activity indexes as well as to see the seismic signature after solar flares (Kumar et al. 2010). Apart from a small time lag between 3 to 5 days, the correlation with the classical activity index was above 50\% reaching in one of the cases $\sim$60\%. Thus, we start by inspecting the light curve of the combined observations in which the presence of starspots is clearly visible (see Top panel of Fig.~1). We use a Global Wavelet Power Spectrum  (GWPS, Torrence \& Compo~1998) to study the energy distribution with time of the surface rotation peak at low frequency corresponding to a period of ~3.4 days as measured by Appourchaux et al. (2008). 

We first confirm the surface rotation rate of $\sim$3.5d with some differential rotation between 2 and 5 days (see middle panels in Fig. 1). If we divide the observations in three regions of 60 days --denoted by A, B, and C-- the GWPS shows a quiet period  --less perturbations due to crossing starspots-- in B while there are much more GWPS power at the beginning of A. This analysis is confirmed by collapsing the GWPS in the frequency direction. A maximum appears at the middle of A (see bottom panel in Fig.~1). 

\begin{figure}[!htbp]
\includegraphics[width=7.5cm, trim = 1cm 2.5cm 2.5cm 3.5cm]{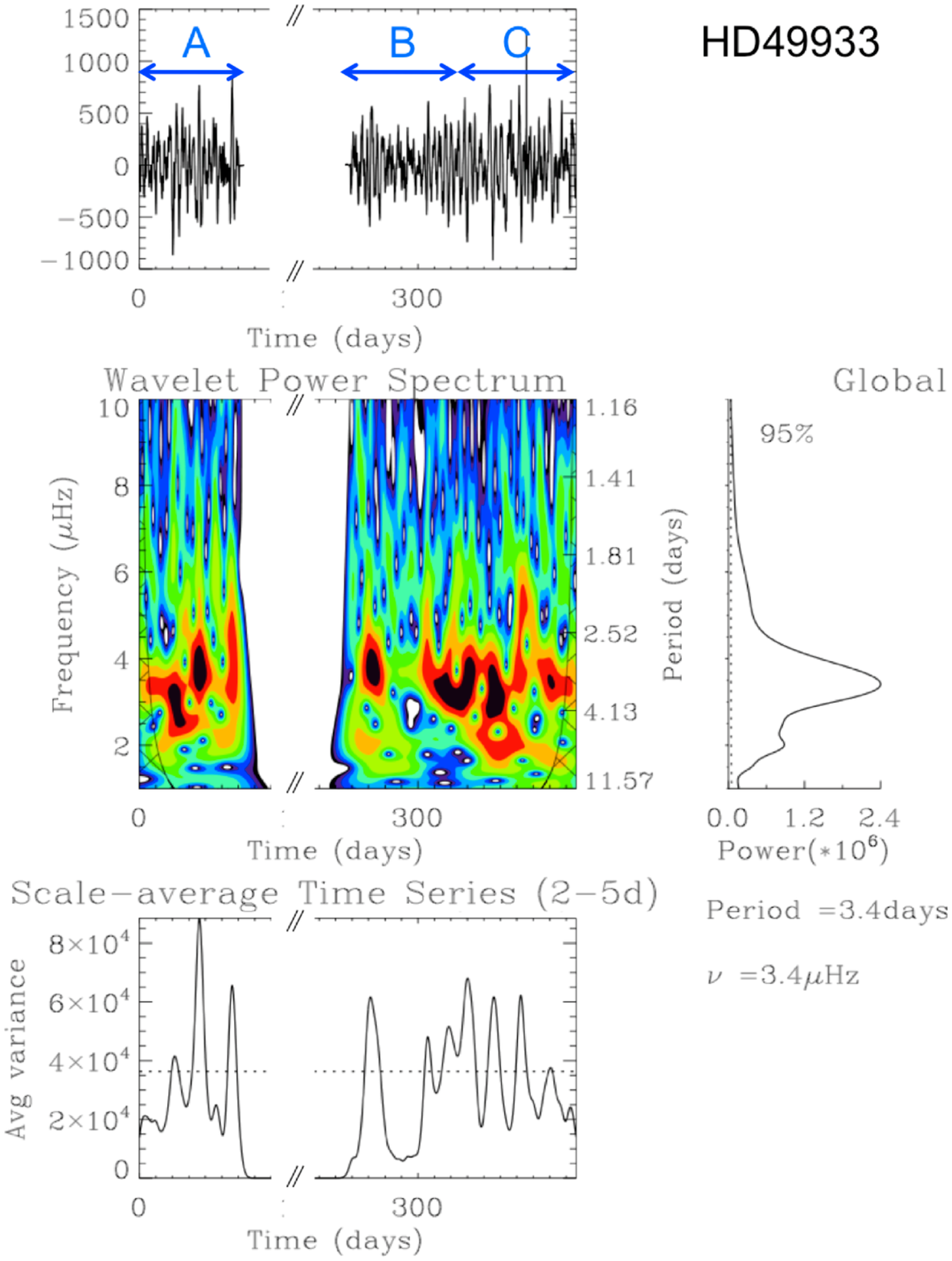}
\caption{Top: Light curve of HD49933. Middle: GWPS and collapsed spectrum in the time direction showing a main surface rotation rate of $\sim$3.5 days. Bottom: Collapsed spectrum in the frequency direction. }
\label{wav}
\end{figure}

\section{Study of the frequency shift}

When the precision in the determination of the frequencies is not high enough, the frequency shift of the p modes can be obtained by analyzing the cross-correlation function between the power spectra of different periods of observation centered on the p-mode hump (Pall\'e, R\'egulo \& Roca Cort\'es, 1989). To do so, three power spectra of 60 days (resolution of 0.19 $\mu$Hz) have been calculated using the two observational runs of HD49933. To reduce the noise, the convective background modeled with a standard Harvey model has been subtracted (e.g. Lefebvre et al. 2008). This background has three components and six free parameters (ignoring the p-modes hump). The cross-correlation function is thus computed in the region of the p-mode hump between 1000 and 2500 $\mu$Hz. 

\begin{figure*}[!htbp]
\includegraphics[angle=-90, width=12cm,trim = 7cm 0.1cm 7cm 9cm]{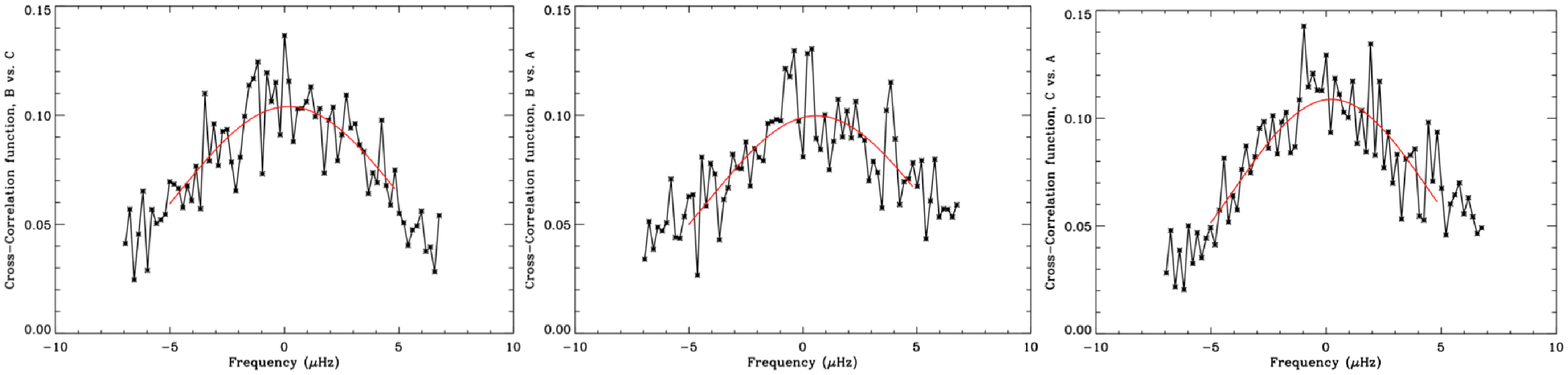}
\caption{Cross-Correlations functions computed between the three 60-days segments A, B and C as described in the text.}
\label{ccorr}
\end{figure*}

Cross-correlation functions (see Fig.~\ref{ccorr}) look like Gaussians. To determine a possible shift in this cross-correlation function, we use a frequency range of ±7$\mu$Hz around the maximum of the cross correlation to estimate the third order moment of this function, which measures the asymmetry. Then, starting with the lag given by the asymmetry, we fit a Gaussian function on a window of $\pm$5$\mu$Hz, that is approximately the second order moment of the cross-correlation function. The maximum of this fitted Gaussian is used as the position of the cross-correlation peak. In the case of solar data, if we use the power spectrum of the minimum as a reference, a positive shift is found.

The result of applying this technique to the three independent segments of 60 days is shown in Fig.~\ref{ccorr}.
The cross- correlating between B and C (the two segments of the second run), gives a maximum of the fitted Gaussian which is shifted $0.17 \pm 0.18$ $\mu$Hz. A positive shift, although it is inside the error bar. The cross-correlation of A with the first 60 days of the long series, B,  produces a shift of $0.59\pm0.23$ $\mu$Hz. Finally, the cross-correlation of A with the second part of the long series, C, shows a shift of $0.23\pm0.16$ $\mu$Hz. 

Although the computed correlations are quite noisy, and following the lessons learnt with the Sun, they could be telling us that a maximum activity occurs during the first short run A, then a minimum arrived during B, and there is an intermediate activity period during C. The same analysis using independent subseries of 30 days gave the same qualitative results but with higher error bars due to the reduction of the time series. 

\section{Time evolution of $A_{max}$}

Finally, we use the A2Z pipeline (Mathur et al. 2010b) to obtain the rms maximum amplitude per radial mode, $A_{\rm{max}}$, of the p-mode hump and its variation with time. To do so, we cut the time series in subseries of 20-day long shifted by 5 days. The convective background is fitted in each subseries with a Harvey-law function as it has been already described to compute the frequency shifts. $A_{\rm{max}}$ is computed by fitting a Gaussian function in the frequency region between 1000 and 2300 $\mu$Hz.

\begin{figure}[!htbp]
\includegraphics[angle=90, width=8cm]{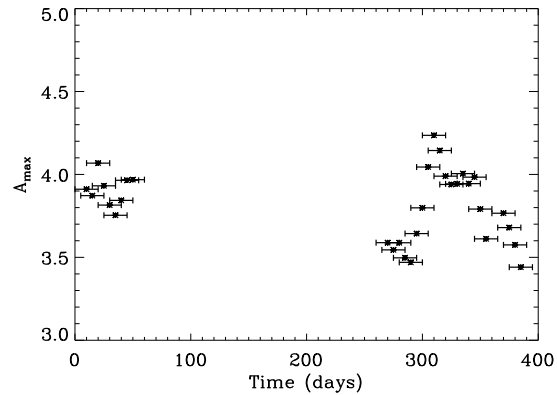}
\caption{Time evolution of the rms maximum amplitude per radial mode. The horizontal error bars represent the range of each subseries of 20 days.}
\label{Amax}
\end{figure}

We found a maximum in $A_{\rm{max}}$ (see Fig.~\ref{Amax}) around the middle of the long run (between B and C). This would indicate a minimum activity, which is in agreement with the frequency shift found when using cross-correlations. However, it is a little shifted in time compared to the quiet period observed in the light curve. 

The amplitude during the first period, A, has an intermediate amplitude while, according to the frequency shifts, it should closer to a minimum amplitude.

\section{Conclusions}
We found a significant shift in the average p-mode frequencies of HD49933 as well as a variation in the maximum amplitude per radial mode with time. Using the Sun as a reference, the frequency-shift found in the analysis is telling us that a period of minimum activity arrives during the first part of the second run.
 A visual inspection of the light-curves also reveals the existence of a magnetic activity cycle because the signature of the perturbations induced by the active regions on the stellar surface changes with time. However, it is not possible to obtain direct inferences on the activity cycle from this observable because the inclination angle of the star should be taken into account as well as some hypothesis on the active longitudes of the star.   
 From the results obtained analyzing the collapsed spectrum in the frequency direction, and the results from the variation of the maximum amplitude per radial mode, this minimum in the activity seems to occur at the end of the first part of the long run.
The period of high activity in the short run unveiled by the analysis of the frequency-shift is confirmed by the collapsed global wavelet spectrum, and from the result of the variation of $A_{\rm{max}}$. Moreover, the frequency shifts and the variation of $A_{\rm{max}}$ are anticorrelated as it is the case in the Sun.
All the analyzed information indicates the existence of a magnetic activity cycle in HD49933. 

\acknowledgements
The CoRoT (Convection, Rotation and planetary Transits) space mission has been developed and is operated by CNES, with the contribution of Austria, Belgium, Brazil, ESA (RSSD and Science Program), Germany and  Spain. S.M. wants to thank the support of the French PNPS.

\end{document}